\documentclass[amsfonts,11pt]{article}
\usepackage{graphicx}
\usepackage{amssymb}
\usepackage{color}

\title{Solutions of Friedmann's Equations\\ and Cosmological Consequences\footnote{This is a survey article based on talks at ICCM 7, Academia Sinica, Beijing, August 6--11, 2016, and International Workshop on Nonlinear Partial Differential Equations and Applications, New York University - Shanghai
and China Eastern Normal University, Shanghai,  June 13--15, 2016.}}
\author{Yisong Yang\\New York University and Henan University}
\date{}

\newcommand{\lm}{\lambda}

\newtheorem{oldtheorem}{Theorem}[section]
\newtheorem{oldassertion}[oldtheorem]{Assertion}
\newtheorem{oldproposition}[oldtheorem]{Proposition}

\newtheorem{oldlemma}[oldtheorem]{Lemma}
\newtheorem{olddefinition}[oldtheorem]{Definition}
\newtheorem{oldclaim}[oldtheorem]{Claim}
\newtheorem{oldcorollary}[oldtheorem]{Corollary}
\newtheorem{thm}{Theorem}[section]

\newbox\qedbox

\newcommand{\dd}{\mbox{d}}
\newcommand{\ee}{\end{equation}}
\newcommand{\be}{\begin{equation}}\newcommand{\bea}{\begin{eqnarray}}
\newcommand{\eea}{\end{eqnarray}}
\newcommand{\e}{\mbox{e}}
\newcommand{\pa}{\partial}

\newcommand{\nn}{\nonumber}

\begin{document}

\maketitle

\begin{abstract}
The Einstein equations of general relativity reduce,
when the spacetime metric is of the
Friedmann--Lema\^{i}tre--Robertson--Walker type governing an isotropic and homogeneous universe, to
the Friedmann equations, which is a set of nonlinear
ordinary differential equations, determining the law of evolution of the
spatial scale factor, in terms of the Hubble ``constant''.
 It is a challenging task, not always possible, to
solve these equations. In this talk, we present some insights from
solving and analyzing the Friedmann equations and their implications to evolutionary cosmology.
In particular, in the Chaplygin fluid universe, we derive
a universal formula for the asymptotic exponential growth rate of the scale factor
which indicates that, as far as there is a tiny presence of nonlinear (exotic) matter, linear (conventional) matter makes contribution to the dark energy, which becomes significant near
the phantom divide line. Joint work with Shouxin Chen, Gary W. Gibbons, and Yijun Li.

\medskip


\medskip

\medskip

\end{abstract}

\section{Introduction}
\setcounter{equation}{0}
\setcounter{figure}{0}

When the spacetime metric is of the
Friedmann--Lema\^{i}tre--Robertson--Walker type governing an isotropic and homogeneous universe filled with a perfect fluid
characterized by its pressure $P_m$ and energy density $\rho_m$,
the Einstein equations of general relativity reduce
  to
the  Friedmann equations which is a set of nonlinear 
ordinary differential equations that determines the law of evolution of the
spatial scale factor $a(t)$  in terms of the  Hubble ``constant'',
\be\label{1.1}
H= \frac{\dot a}{a}.
\ee
It has been a challenging task to
obtain a complete explicit integration of these equation in finite terms in order to gain a precise understanding of the evolution of the universe in view of
the scale factor.
In a series of recent work \cite{CGLY,CGY1,CGY2}, S. Chen, G. Gibbons, Y. Li, and the author succeeded in obtaining considerable systematic insight
into these difficult nonlinear ordinary differential equations. Specifically, in \cite{CGLY}, we demonstrate a link between  
a theorem of  Chebyshev   and the explicit integration
in both cosmological time $t$ and conformal time $\eta$ of the Friedmann equations
in all dimensions and with an arbitrary cosmological constant $\Lambda$.
We are able to establish that for spatially  flat universes  an explicit
integration
in $t$ may always be carried out, and that,
in the non-flat situation and when $\Lambda$ is zero
and
the ratio $A$ of $P_m$ and $\rho_m$ in the barotropic equation of state,
\be\label{1.2}
P_m=A\rho_m,
\ee
 of the perfect-fluid universe is rational,  an explicit integration may be carried out if and only if the dimension $n$ of space and $A$ obey some specific relations among an infinite family. Besides, we proved that the situation for an explicit integration in $\eta$ is complementary to
that in $t$ so that in the flat-universe case
with $\Lambda\neq0$ an explicit integration in $\eta$  can
be carried out if and only if $A$ and $n$ obey similar
relations among a well-defined  family,
and that, when $\Lambda=0$, an explicit
 integration can always be carried out whether the
space is flat, closed, or open.  In \cite{CGY1}, we explore the method further to get exact integrations of
some other important cases when the equation of state of the perfect-fluid universe is of a nonlinear type,
\be\label{1.3}
P_m=f(\rho_m),
\ee
 and settle some important cases whose integrability is beyond the reach of
the Chebyshev theorem but can be obtained by other means of integration, including
 the generalized Chaplygin gas \cite{ST,CP,ABL}, 
two-term energy density \cite{NP1,NP2}, the trinomial Friedmann \cite{AC},
Born--Infeld, and two-fluid models. In particular, for the
generalized Chaplygin gas model, we work on the flat-universe situation with zero cosmological constant and identify all integrable cases. For the two-term
energy density model, we can do the same. In both situations the Chebyshev theorem may be applied directly. In the trinomial Friedmann equation model, we show
that the Chebyshev theorem is applicable only in the bottom three-dimensional space case. For the Born--Infeld type fluid model and assuming a flat universe, we show that the Chebyshev theorem works only in the critical coupling situation when the cosmological constant and Newton constant fulfill a specified condition. However, in all non-critical cases, we show that the Friedmann equation allows an integration for any value of the cosmological constant.
Moreover, we study a two-fluid model and carry out its integration by the Chebyshev theorem for a closed universe situation. Its interest is that in the original
setting the integrand is not of a binomial type but it may be recast and decomposed into a binomial form so that the Chebyshev theorem is applicable. 
We will also conduct a study of the two-fluid model in terms of the reduced temperature, given as the inverse of the scale factor. In this situation the Friedmann
equation in conformal time does not allow an application of the Chebyshev theorem because it cannot be reduced into a binomial form but may still be integrated
explicitly when the cosmological constant vanishes and universe is either closed or open (the flat case is trivial). The necessity of introducing several layers of
intermediate variables in the process of integration often makes it complicated and cumbersome to express the 
 final dependence of the scale factor on either
cosmic or conformal time transparently, although such a task is always manageable. As  illustrations, we have worked out 
concrete examples for the models which are integrated in terms of cosmic time. These include the Chaplygin gas, Born--Infeld, two-term energy, two-fluid, and
trinomial Friedmann equation models. Although these examples are of different technical features, the common lesson gained is that  integration allows us to
obtain both qualitatively and quantitatively accurate knowledge about the solutions, especially regarding the roles played by various physical parameters.
For example, in the Chaplygin gas universe, explicit formulas are obtained for the exponential expanding rates of the scale factor which indicate that, no matter how
small the nonlinear Chaplygin component is, the linear component always join force to contribute to the exponential rate, or dark matter, even though the cosmological
constant is absent.
Moreover, near the so-called phantom divide line, the linear component contributes to dark matter significantly.
In \cite{CGY2}, we present
several analytic methods which may be used to
obtain either exact expressions or insightful knowledge of the solutions of the Friedmann equations
beyond the reach of the Chebyshev theorem. These methods are divided into three categories:
roulettes, explicit integrations, and analytic approximations.
Note that the Chebyshev theorem applies only to  integrals of binomial differentials. In cosmology, one frequently encounters models which cannot be converted into
such integrals. Hence it will be useful
to present some methods which may be used
to integrate the Friedmann equations whose integrations
involve non-binomial differentials.
There we illustrate our methods by integrating the model when the equation of state is quadratic
\cite{AB1,AB2,SSC}
and the Randall--Sundrum II universe \cite{Ge,RS,SWC,CC} with a non-vanishing cosmological constant for which both the energy density and its quadratic power are contributing to the right-hand side of the Friedmann equation. A quadratic equation of state introduces a larger degree of freedom for the choice of
parameters realized as the coefficients of the quadratic function. In
the flat-space case we show that for the solutions
of cosmological interest to exist so that the scale factor evolves from zero to infinity the coefficients
must satisfy a  necessary and sufficient condition that confines the ranges of the parameters. We also
show that when  the quadratic term is present in
the equation of state the  scale factor cannot vanish at finite time. In other
words, in this situation, it requires an infinite past duration
for the scale factor to vanish.
As another example that cannot be dealt with by the Chebyshev theorem, we study the Randall--Sundrum
II universe when the space is flat and $n$-dimensional and the cosmological constant $\Lambda$ is arbitrary. There are two cases of interest: (i) the equation of state is linear, and,
(ii) the equation of state is that of the Chaplygin fluid type.
In the case (i) the big-bang solutions for $\Lambda=0$ and $\Lambda>0$ are similar as in the classical situation so that the scale factor either enjoys a power-function growth law or an exponential growth law according to whether $\Lambda=0$ or $\Lambda>0$. When $\Lambda<0$, however, we unveil the
fact that the solution has only a finite lifespan when the coupling parameters lie in some explicitly stated ranges , and we determine the
lifespan also explicitly. Otherwise, the solution is periodic, as in the classical case.
In the case (ii) with $\Lambda=0$, we find the explicit big-bang solution by
integration and display its
exponential growth law in terms of various physical constants in the model. Besides, we have carried
out a study of the big-bang solution for the Friedmann equation when the equation of state is of
an extended Chaplygin fluid form \cite{PK1,PK2,KKPMP}, 
\be
P_m=f(\rho_m)-\sum_{k=1}^m\frac{B_k}{\rho_m^{\alpha_k}}, 
\ee
where
$f(\cdot)$ is an analytic function and $B_1,\dots,B_m>0,\alpha_1,\dots,\alpha_m\geq0$, for which
an integration by whatsoever means would be impossible in general. We can identify an
explicit range of $\Lambda$ for which the scale factor grows exponentially fast and deduce a universal formula for
the associated exponential growth rate. This formula covers all the explicitly known concrete cases.

After such an extensive description of the results of \cite{CGLY,CGY1,CGY2}, below we aim to report and highlight concisely some of these in a popular-science talk manner.

For some earlier works on the integration of the Friedmann equations, we refer the reader to
\cite{Barrow,Gibbons,SK,F,L,UL,Linder,CDS,van,W,Harrison,ST,WKS,FS0,CS,sc1} and references therein, as well as the bibliography cited in \cite{CGLY,CGY1,CGY2}.

\section{Einstein equations and  consequences}
\setcounter{equation}{0}
\setcounter{figure}{0}

The Einstein equations in an $(n+1)$-dimensional spacetime are written
\be
G_{\mu\nu}+\Lambda g_{\mu\nu}=8\pi G_n T_{\mu\nu},\quad \mu,\nu=0,1,\dots,n,
\ee
where $G_{\mu\nu}$ is the Einstein tensor induced from the spacetime metric tensor $g_{\mu\nu}$, $G_n>0$ the universal gravitational constant in $n\geq3$ spatial dimensions, $\Lambda$ the cosmological constant, and $T_{\mu\nu}$ the energy-momentum tensor of the matter in the universe.

Even when the energy-momentum tensor is regarded as given, these equations govern
\be
1+2+\dots+(n+1)=\frac12 (n+1)(n+2)
\ee
unknowns, $g_{\mu\nu}$,  and are highly nonlinear and difficult to understand.

Two over-simplified cases though are manageable and have provided enormous insight toward an understanding of our universe based on the Einstein equations. They are
\begin{enumerate}
\item[(i)] Time-independent solutions with spherical symmetry. These solutions have given births to many important concepts such as
black holes and their formation, spacetime singularities,
gravitational collapse, black-hole evaporation, and spacetime entropy.
The names of the main contributors include K. Schwarzschild, A. Eddington, R. Kerr, E. Newman, D. Finkelstein, R. Penrose, S. Hawking, G. Gibbons, J. Bekenstein.

\item[(ii)] Time-dependent spatially-uniform solutions. These solutions led to the conceptualization of  various modern thoughts on our universe such as
big bang cosmology, cosmic inflation and radiation, and dark energy.
The names of the main contributors in this area include A. Friedmann,  R. Tolman, E. Hubble, A. Eddington,  S. Hawking, A. Guth, A. Linde, A. Starobinsky.

\end{enumerate}

Here we are  interested in the subject (ii).

Recall that, according to Cosmological Principle, the universe is homogeneous and isotropic in large scale. Such an assumption
implies that there is a simultaneous
 cosmic time, $t$, and that the space curvature is constant
for fixed time.
As a consequence the metric element takes the form
\be
\dd s^2=-\dd t^2+a^2(t)\left(\frac1{1-k r^2}\dd r^2+r^2\dd\Omega_{n-1}^2\right),
\ee
known as the Friedmann--Lema\^{i}tre--Robertson--Walker, or Robertson--Walker,  metric, where $k=0,\pm1$ is an indicator such that, when
$k=1$ the curvature is positive and the space is referred to as closed, $k=-1$ the curvature is negative and
the space is open, and  $k=0$ the curvature is zero and the space is flat,
the speed of light is taken to be unit, $c=1$, $\dd\Omega^2_{n-1}$ is the standard metric element of $S^{n-1}$,
$a(t)>0$ is the scale factor, nicknamed ``radius of universe'', and
\be
K(t)=\frac1{a^2(t)}
\ee
is the space curvature when $k=1$ and the space is actually realized as
\be
x_1^2+\cdots +x_{n+1}^2=a^2(t), 
\ee
which is an $n$-sphere of radius $a(t)$.

Some of the prominent physical features of the scale factor are as follows.

\begin{enumerate}
\item[(i)] Temperature

Assume that the cosmological particles are in thermodynamic equilibrium. Then the absolute temperature of the universe is \cite{Coq}
\be
T=\frac{C_0} a,
\ee
where $C_0>0$ is a constant.
So at the ``big bang" epoch, $a\sim 0$, the universe is extremely hot and cooling happens as the universe is expanding, $a\gg1$.

\item[(ii)] Redshift

Use $\lm$ to denote the wavelength. Then the redshift $z=\frac{\lm-\lm_0}{\lm_0}$ in terms of $a$ obeys the law
\be
1+z=\frac{a(t_0)}{a(t)},\quad t>t_0.
\ee
So  the expansion of the universe leads to the observation of a redshift.

\item[(iii)] Retreating speed

Use $R$ to denote the distance between two remote stars or galaxies. Then the ``retreating speed", $\dot{R}$, measures how fast the stars
or galaxies recede from each other and
is proportional to the Hubble constant, given in terms of the scale factor, following the formula
\be
\frac{\dd R}{\dd t}=\frac{\dot{a}}{a}R.
\ee
\end{enumerate}

Thus, measurement of any of the above three quantities would offer clues on how the universe expands.

Assume now that ours is a perfect-fluid universe. Then the energy-momentum tensor is totally determined by two quantities, $\rho_m$ and $P_m$, where
$\rho_m$ is the time-dependent-only rest-frame mass (energy) density and $P_m$ the isotropic pressure, so that the mixed energy-momentum tensor of the cosmological fluid reads
\be
T^\nu_\mu=\mbox{diag}\left\{-\rho_m,P_m,\dots,P_m\right\}.
\ee

With the Robertson--Walk metric and the above perfect-fluid energy-momentum tensor, we see that the Einstein equations are reduced into the following Friedmann equations
\bea
H^2&=&\frac{16\pi G_n}{n(n-1)}\,\rho-\frac k{a^2},\label{2.10}\\
 \dot{H}&=&-\frac{8\pi G_n}{n-1}(\rho+P)+\frac k{a^2},\label{2.11}
\eea
where $H$ is the Hubble parameter given in (\ref{1.1}) and
\be\label{2.12}
\rho=\rho_m+\frac{\Lambda}{8\pi G_n},\quad P=P_m -\frac{\Lambda}{8\pi G_n},
\ee
are the effective energy density and pressure, which are subject to the
law of energy conservation,
\be\label{2.13}
\dot{\rho}+n(\rho+P)H=\dot{\rho}_m+n(\rho_m+P_m)H=0.
\ee

The barotropic equation of state of the fluid is a statement that $P_m$ is a function of $\rho_m$,
of the form (\ref{1.3}), whose simplest form is linear, as stated in (\ref{1.2}), for which the two classical conventional situations are
(i) matter (dust) dominated, $A=0$, and
(ii) radiation dominated, $A=\frac1n$.

\begin{figure}
\begin{center}
\includegraphics[height=6.2cm,width=11cm]{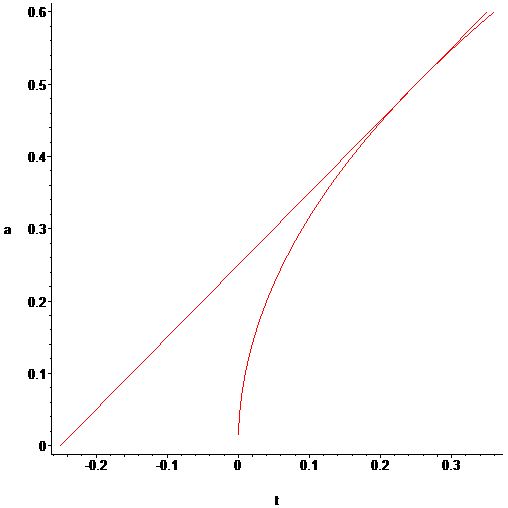}
\caption{Evolution of the scale factor $a(t)$ versus the cosmic time $t$. Concavity of $a$ indicates the existence of a finite past time when $a$ vanishes.}
\label{Fig1}
\end{center}
\end{figure}

To proceed, we begin by considering the vanishing cosmological constant case, $\Lambda=0$.
The Friedmann equations (\ref{2.10})--(\ref{2.11}) may be combined to give us
\be
\frac{\ddot{a}}a=-\frac{8\pi G_n}{n(n-1)} \left([n-2]\rho_m +n P_m\right).
\ee
Thus, in both classical cases, we have the concavity property
$
\ddot{a}(t)\leq0.
$
However,
Hubble's result indicates
\be
H(t_0)=\frac{\dot{a}(t_0)}{a(t_0)}>0,\quad t_0=\mbox{``today"}.
\ee
Hence the universe must have a finite past when $a=0$, inevitably leading to the so-called
``Big Bang'' scenario, as shown in Figure \ref{Fig1}.

Furthermore,  it may be checked that (\ref{2.10}) implies (\ref{2.11}), subject to (\ref{2.13}). Thus it suffices to
consider the first Friedmann equation, (\ref{2.10}),
\be
H^2+\frac k{a^2}=\frac{16\pi G_n}{n(n-1)}\,\rho,
\ee
also known as the Raychaudhuri equation.
Let us specialize on the flat case $k=0$. Then Hubble's result $H(t_0)>0$ indicates that
$
H(t)>0$ for all  $t$, past and future.
Therefore the universe has been and will be expanding.

We now consider the simplest situation where $\Lambda=0$ and $k=0$ and the equation of state is linear as given in (\ref{1.2}).
Inserting this into the law of energy conservation, (\ref{2.13}),
we obtain
\be
\rho_m =\rho_0 a^{-n(1+A)},
\ee
which clearly indicates that the energy density decreases as the universe expands (when $A>-1$).

Using the above in the Raychaudhuri equation we arrive, via solving
\be
\left(\frac{\dot{a}}a\right)^2=\frac{16\pi G_n}{n(n-1)}\rho_0 a^{-n(1+A)},
\ee
 at the well-known solutions:
\bea
a(t)&=&C_0 t^{\frac2{n(1+A)}},\quad  A>-1, \quad a(0)=0, \quad t>0,\nn\\
a(t)&=& a(0) \e^{4\sqrt{\frac{\pi G_n\rho_0}{n(n-1)}}t}, \quad A=-1,\quad a(0)>0, \quad t>0,\nn\\
a(t)&=&
\left(a^{\frac{n(1+A)}2}(0)-4\left[\frac{\pi G_n \rho_0}{n(n-1)}\right]^{\frac12} t\right)^{\frac2{n(1+A)}},\quad
 A<-1,\quad a(0)>0,\quad t>0,\nn
\eea
which prompt us to describe them as follows.
\begin{enumerate}
\item[(i)] $A>-1$. This is the conventional situation with big bang.

\item[(ii)] $A<-1$. This is a non-conventional (``phantom" or ``ghost") situation \cite{Ho,CK} without big bang (since $a(0)>0$) and with finite  life time $t=T_0>0$ given as
\be
T_0=\frac{a^{\frac{n(1+A)}2}(0)}{4\left(\frac{\pi G_n\rho_0}{n(n-1)}\right)^{\frac12}},
\ee
such that
$
\dot{a}(T_0)=\infty$ and $ \ddot{a}(T_0)=\infty$.
That is, the universe blows up at infinite speed and acceleration. Such a situation is known as the Big Rip \cite{EMM,CKW}.

\item[(iii)] $A=-1$. This is the borderline, between the conventional and non-conventional domains, called the ``phantom divide line" or PDL \cite{Mc,C-L,N-P,G-C}.

\end{enumerate}

Summarizing the above, we can draw the following observations.
\begin{enumerate}
\item[(i)] Conventional matter does not lead to exponential growth.

\item[(ii)] Phantom matter leads to the big rip.

\item[(iii)] At the critical case $A=-1$ or PDL, we have the exponential growth but cannot have the big bang state, $a(0)=0$.

\end{enumerate}

\section{Dark energy in terms of scale factor}
\setcounter{equation}{0}
\setcounter{figure}{0}

\begin{figure}
\begin{center}
\includegraphics[height=5.2cm,width=8cm]{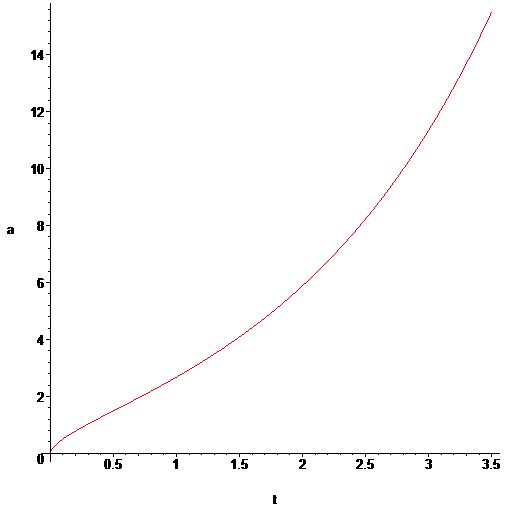}
\caption{Evolution of the universe in terms of the scale factor in presence of dark energy. In the far future the scale factor $a(t)$ would grow acceleratingly fast following an exponential growth
pattern.}
\label{Fig2}
\end{center}
\end{figure}

Astronomical observations over the last two decades have established the fact that the universe is expanding at an accelerating rate. Expressing such a
consensus in terms of the evolution pattern of the scale factor $a(t)$, we would have $\ddot{a}(t)>0$, rather than $\ddot{a}(t)\leq0$, of a decelerating manner.
In modern cosmology, dark energy is an unknown, hypothetical, form of energy assumed to occupy the full universe and responsible for the accelerated  expansion of the universe. Figure \ref{Fig2} depicts an expected growth pattern of the scale factor in presence of dark energy. Mathematically, such a growth pattern spells out
the 
exponential growth law 
\be\label{3.1}
a(t)\sim \e^{\alpha t},\quad \alpha>0,\quad t \to\infty,
\ee
that a correct theory should contain, which is our present main interest.

In literature the exponential growth rate $\alpha$ in (\ref{3.1}) is often simply referred to as
``dark energy" which we also adopt here.

We explore several scenarios for the presence of dark energy.
\begin{enumerate}
\item[(i)] The first scenario is through the presence of a cosmological constant $\Lambda$. 
 We assume conventional matter, $A>-1$, and flatness, $k=0$.

If $\Lambda>0$, we have
\be\label{3.2}
a^{n(1+A)}(t)=\frac{8\pi G_n\rho_0}{\Lambda}\sinh^2\left(\sqrt{\frac{n\Lambda}{2(n-1)}}(1+A) t\right),\quad t>0.
\ee
Hence big bang, $a(0)=0$, and dark energy are both present. Physical interpretation: In
view of (\ref{2.12}), which provides the effective energy density and pressure,
we see that the presence of a positive $\Lambda$ adds an energetic background for the universe to behave more ``agitatedly" and at the same time renders a negative pressure to drag the universe
apart.

If $\Lambda<0$, we have
\be\label{3.3}
a^{n(1+A)}(t)=\frac{8\pi G_n\rho_0}{(-\Lambda)}\sin^2\left(\sqrt{\frac{n(-\Lambda)}{2(n-1)}}(1+A) t\right), \quad t>0.
\ee
Hence we have the big bang and a periodic universe, also known as a big bang - big crunch universe. Physical interpretation:
Again in view of (\ref{2.12}), in a parallel manner, the universe now stays in a less energetic background and the enhanced pressure works to hold the universe together. As a result, it oscillates periodically.

In both cases it will be interesting to describe the dependence of dark energy when $\Lambda>0$ and oscillation period when $\Lambda<0$ with regard to the value of the coupling parameter $A$, especially
when it is in the vicinity of PDL. We leave this to the reader as an exercise.

\item[(ii)] The second scenario is centered around a unified action principle of the form
\be
S=\int\left\{\frac R{16 \pi G_n}+{\cal L}\right\}\sqrt{|g|}\,\dd x,
\ee
combining the internal Einstein--Hilbert action and an external matter-field action for which the matter fields, called the inflation fields, are used to propel
the exponential growth of the universe. In the context of such a formalism there are investigations based on ideas of quintessence
\cite{Sal,Linde0,Linde1,CYam}, the Brans--Dicke theory \cite{DB,CH}, tchyons \cite{GK,Sen}, etc.

\item[(iii)] The third scenario is to simulate the universe by appropriate perfect-fluid models based on various nonlinear equations of state, (\ref{1.3}), which
is the focus of our studies \cite{CGLY,CGY1,CGY2}. 
\end{enumerate}

\section{Insight from Chebyshev's theorem}
\setcounter{equation}{0}
\setcounter{figure}{0}

For simplicity we begin by considering the linear equation of state situation and we work on the Friedmann equation with $\Lambda=0$ and $k\neq0$,
\be
\dot{a}^2=\frac{16\pi G_n\rho_0}{n(n-1)} a^{-n(1+A)+2}-k,
\ee
which leads to the integration
\bea\label{4.2}
I&=&\int a^{\frac12n(1+A)-1}\left(-k a^{n(1+A)-2}+\sigma\right)^{-\frac12}\,\dd a\nn\\
&=&\frac2{n(1+A)}\int\left(-k u^\gamma+\sigma\right)^{-\frac12}\,\dd u,\\
 u&=&a^{\frac12n(1+A)},\quad \gamma=2\left(1-\frac2{n(1+A)}\right),\quad \sigma=\frac{16\pi G_n\rho_0}{n(n-1)}.\nn
\eea

Since (\ref{4.2}) is of a binomial type,  we are prompted to recall the Chebyshev theorem \cite{C0,C1} as follows.

\begin{thm}[Chebyshev's Theorem]\label{thm4.1}
Let $p,q,r$ $ (r\neq0)$ be rational numbers and $a,b$ two nonzero real numbers. Then
\be
\int x^p (a+b x^r)^q\,\dd x
\ee
may be integrated in terms of elementary functions if and only if at least one of the fractions
\be
\frac{p+1}r,\quad q, \quad\frac{p+1}r +q,
\ee
is an integer.
\end{thm}

In view of the Chebyshev theorem we see that when the coupling parameter $A$ in the equation of state is rational the integral (\ref{4.2})  is elementary if and only if
\be
\frac1\gamma\quad \mbox{or}\quad\frac{2-\gamma}{2\gamma}
\ee
is an integer (the trivial case $\gamma=0$ is excluded).
Hence we obtain all solvable values of $A$ as follows:
\bea
A&=&\frac{4N}{n(2N-1)}-1,\quad N=0,\pm1,\pm2,\dots,\label{4.6}\\
A&=&\frac2n+\frac1{nN}-1,\quad N=\pm1,\pm2,\dots.\label{4.7}
\eea

In particular, in the special situation when $n=3$, we have
\be
A=-1,-\frac23,-\frac59,-\frac12,-\frac7{15},-\frac49,-\frac37,\dots,-\frac29,-\frac15,-\frac16,-\frac19,0,\frac13,
\ee
in which $A=0$ (dust) and $A=\frac13$ (radiation) are well known \cite{SK}.

It will also be enlightening to turn to the conformal time $\eta$ with
$
\dd \eta=\frac{\dd t}a
$
and the notation
$
a'=\frac{\dd a}{\dd\eta}
$.
In this context we see that the Friedmann equation with $\Lambda\neq0$ and $k=0$ becomes
\be
(a')^2=\frac{16\pi G_n\rho_0}{n(n-1)} a^{-n(1+A)+4}+\frac{2\Lambda}{n(n-1)} a^4,
\ee
whose integration is elementary if and only if
\be
\frac1{n(1+A)}\quad\mbox{or}\quad \frac1{n(1+A)}-\frac12
\ee
is an integer (assuming $A\neq -1$ which is the PDL).
More precisely, we obtain the following solvable cases:
\bea
A&=&-1+\frac1{nN},\quad N=\pm1,\pm2,\dots,\\
A&=&-1+\frac1{n\left(N+\frac12\right)},\quad N=0,\pm1,\pm2,\dots.
\eea
It is interesting to note that the PDL, $A=-1$, stands out in the large $|N|$ limit.

On the other hand, however, in cosmic time we have the limit
\be\label{4.13}
\lim_{|N|\to\infty} A=\frac2n -1,
\ee
for both cases (\ref{4.6}) and (\ref{4.7}). Unlike in conformal time,  this latter  result,  (\ref{4.13}),  seems to have little classical significance.

\section{Chaplygin fluid universe}
\setcounter{equation}{0}
\setcounter{figure}{0}

In a Chaplygin fluid universe the equation of state is
\be\label{5.1}
P_m=A\rho_m -\frac B{\rho_m^\alpha},\quad 0\leq\alpha\leq1, \quad A,B>0,
\ee
where $A,B,\alpha$ are constants.
In order to stay within the conventional fluid situation in the $B\to0$ limit, we will assume 
\be
A>-1.
\ee

Inserting (\ref{5.1}) into the law of energy conservation (\ref{2.13}),
we find
\be\label{5.3}
(1+A)\rho_m^{\alpha+1}=C a^{-n(1+A)(\alpha+1)} +B,\quad C>0.
\ee
Inserting (\ref{5.3})  into the Friedmann equation (\ref{2.10}), with $k=0$ and $\Lambda=0$,
we have
\be\label{5.4}
\dot{a}^2=\frac{16\pi G_n}{n(n-1)} (1+A)^{-\frac1{\alpha+1}} a^2\left(Ca^{-n(1+A)(\alpha+1) }+B\right)^{\frac1{\alpha+1}},
\ee
which prompts to the integration
\be\label{5.5}
I=\int\left(C a^{-n(1+A)(\alpha+1)} +B\right)^{-\frac1{2(\alpha+1)}}\frac{\dd a}{a}.
\ee
Using Theorem \ref{thm4.1} we see that (\ref{5.5}) allows integration provided that $\alpha$ is rational. It can be seen that
in all integrable cases the equation has the desired solution with the initial condition $a(0)=0$ (big bang) and asymptotic behavior
$a(t)\sim \e^{\lm t}$ for $t\gg1$ (dark energy). Below are some explicit values of the exponential growth rate $\lm$ in terms of $\alpha$:
\bea
\lm&=&4\left(\frac{\pi G_n}{n(n-1)}\right)^{\frac12}\left(\frac B{1+A}\right)^{\frac14},\quad\alpha=1,\label{5.6}\\
\lm&=&4\left(\frac{\pi G_n}{n(n-1)}\right)^{\frac12}\left(\frac B{1+A}\right)^{\frac13},\quad\alpha=\frac12,\label{5.7}\\
\lm&=&4\left(\frac{\pi G_n}{n(n-1)}\right)^{\frac12}\left(\frac B{1+A}\right)^{\frac38},\quad\alpha=\frac13.\label{5.8}
\eea

These results indicate a clear pattern in the exponential growth rate $\lm$:

\begin{enumerate}
\item[]
It contains a common factor which depends only on the spatial dimension $n$ and the Newton constant $G_n$.

\item[]
It is proportional to some power $\gamma$ of the common ratio
$
\frac B{1+A}
$
composed from the linear (conventional) and nonlinear (non-conventional) matter coupling constants $A$ and $B$ in the equation of state of the Chaplygin fluid so that
\be
\gamma=\frac14,\frac13,\frac38,\quad\mbox{when}\quad \alpha=1,\frac12,\frac13,
\ee
respectively.
\end{enumerate}

Suggested by the above results, we are interested in obtaining a growth rate formula for all values of $\alpha$. To this end, we 
use the substitution
\be
w=\left(\frac CB a^{-n(1+A)(\alpha+1)}+1\right)^{\frac1{2(\alpha+1)}},
\ee
and the boundary condition
\be
w(0)=\infty,\quad w(\infty)=1,\quad\mbox{derived from }a(0)=0,\quad a(\infty)=\infty,
\ee
to carry out an analytic investigation. We have obtained the following general formula for $\lm$:
\be\label{5.12}
\lm=4\left(\frac{\pi G_n}{n(n-1)}\right)^{\frac12}\left(\frac B{1+A}\right)^{\frac1{2(\alpha+1)}},\quad \alpha\in[0,1],
\ee
which reveals a {\bf universal formula} for the exponential growth rate for the Chaplygin fluid universe. It is seen that (\ref{5.6})--(\ref{5.8}) are covered as
special cases.

Two immediate implications of the universal growth law are:

\begin{enumerate}
\item[(i)]
Although the linear (conventional) matter ($A>-1$) alone does not give rise to exponential growth, it plays a role whenever a nonlinear (non-conventional) matter is present.

\item[(ii)]
No matter how weak the nonlinear term is, its presence switches on the exponential expansion of the universe which becomes significant near the
PDL limit,
$
A=-1.
$
\end{enumerate}

Thus we conclude that dark energy is present in a flat Chaplygin fluid universe without a cosmological constant.

\section{Chern--Simons modified gravity theory}
\setcounter{equation}{0}
\setcounter{figure}{0}

This four-spacetime dimensional theory is proposed by Jackiw and Pi \cite{JP}. Here we follow their formalism.

Use $^*RR$ to denote the Pontryagin density
\be
^* RR=\,^*R^{\alpha\mu\nu}_\beta R^\beta_{\alpha\mu\nu},
\ee
where $R^\beta_{\alpha\mu\nu}$ is the Riemann tensor and 
\be
^*R^{\alpha\mu\nu}_\beta=\frac12\epsilon^{\mu\nu\mu'\nu'}R^\alpha_{\beta\mu'\nu'},
\ee
 the dual tensor. Then $^*RR$ may be represented as a total divergence \cite{JP,GY,FMN}:
 \be
 \frac14\, ^*RR=\nabla_\mu K^\mu,
 \ee
 with
 \be
 K^\mu=\epsilon^{\mu\nu\alpha\beta}\left(\Gamma^{\nu'}_{\nu\mu'}\pa_\alpha \Gamma^{\mu'}_{\beta\nu'}+\frac23 \Gamma^{\nu'}_{\nu\mu'}\Gamma^{\mu'}_{\alpha\alpha'}\Gamma^{\alpha'}_{\beta\nu'}\right)
 \ee
resembling a contracted Chern--Simons form, in terms of the usual Christoffel symbols.

The modified Hilbert--Einstein--Chern--Simons action is \cite{FNP}
\be
S=\int\left\{\sqrt{|g|}R+\frac l4\Theta ^*RR-\frac12\sqrt{|g|}\pa^\mu\Theta\pa_\mu\Theta\right\}\,\dd x.
\ee
In the homogeneous and isotropic universe case the field $\Theta$ is governed by the wave equation \cite{FNP}
\be
\ddot{\Theta}+3H\dot{\Theta}=0,
\ee
which leads to the integral $\dot{\Theta}=Ca^{-3}$ and the law of energy conservation
\be
\frac{\dd}{\dd t}\left(\rho_m+\frac{C^2}2 a^{-6}\right)+3H(\rho_m+P_m+C^2 a^{-6})=0,
\ee
which actually simplifies into the usual one without the integration constant $C$:
\be
\dot{\rho}_m+3H(\rho_m+P_m)=0.
\ee

The flat-space ($k=0$) Friedmann equation is
\be
H^2=\frac{8\pi G}3\left(\rho_m +\frac{C^2}2 a^{-6}\right).
\ee
Thus, with the linear equation of state (\ref{1.2})  leading to $\rho_m=\rho_0 a^{-3(1+A)}$, we arrived at the modified Friedmann equation
\be\label{6.10}
\left(\frac{\dot{a}}a\right)^2=\frac{8\pi G}3\left(\rho_0 a^{-3(1+A)}+\frac{C^2}2 a^{-6}\right)\equiv\frac{8\pi G}3\rho_{\mbox{\small{eff}}},
\ee
where
\be
\rho_{\mbox{\small{eff}}}= \rho_0 a^{-3(1+A)}+\frac{C^2}2 a^{-6},
\ee
may be regarded as the effective energy density of a two-fluid model.

We are to integrate the equation (\ref{6.10}).

Integrating (\ref{6.10}) we have 
\be
I\equiv \int a^2\left(\rho_0 a^{3(1-A)}+\frac{C^2}2\right)^{-\frac12}\,\dd a=2\sqrt{\frac{2\pi G}3}\, t.
\ee
From Theorem \ref{thm4.1} we know, when $A$ is rational,  that $I$ is elementary if and only if $A$ assumes the values
\bea
A&=&1-\frac1N,\quad N=\pm1,\pm2,\dots,\\
A&=&1-\frac2{2N+1},\quad N=0,\pm1,\pm2,\dots.
\eea

Example: $N=1$ then $A=0$ (dust) or $A=\frac13$ (radiation) so that
\be
a(t)\sim t^{\frac23},\quad a(t)\sim t^{\frac12},\quad t\to\infty,
\ee
respectively.
 In general, there is an absence of exponential growth, or dark energy,  when $\Lambda=0$, which is parallel to the situation in the classical Einstein theory
\cite{CGY1}.

It will be an interesting future project to consider $\Lambda=0$ continuously and choose the equation of state (\ref{1.3})
beyond that of a linear relation, (\ref{1.2}), aiming at  achieving  exponential growth in the Chern--Simons modified
gravity theory.

\section{Born--Infeld theory}
\setcounter{equation}{0}
\setcounter{figure}{0}

Among the simplest nonlinear models is the Born--Infeld model
whose equation of state reads
\be\label{7.1}
P_m=\frac{\rho_m}{\rho_m+1}.
\ee
Now set
$
\tilde{\rho}_m=\rho_m+1$ and $ \tilde{P}_m=P_m-1$.
Then we obtain the updated equation of state:
\be\label{7.2}
\tilde{P}_m=-\frac1{\tilde{\rho}_m},
\ee
which is of a Chaplygin fluid type and might lead to exponential growth.
Here, however, we show that this is not the case as anticipated.

From inserting the equation of state and solving the law of energy conservation we get
\be
\rho_m=-1+\sqrt{1+C_0 a^{-2n}},\quad C_0>0.
\ee
Then the  flat-space Friedmann equation reads
\be
H^2=\frac{16\pi G_n}{n(n-1)}\left(-1+\sqrt{1+C_0 a^{-2n}}+\frac{\Lambda}{8\pi G_n}\right).
\ee
In order to be able to utilize the Chebyshev theorem one needs to restrict to
the critical case
$
\Lambda=8\pi G_n$.
Here we show how to overcome this difficulty.
We set
\be\label{7.5}
\alpha=\frac{\Lambda}{8\pi G_n}-1,\quad\beta^2=\frac{16\pi G_n}{n(n-1)}.
\ee
Then we arrive at the Friedmann equation
\be
(\dot{a})^2=\beta^2 a^2 \left(\alpha+\sqrt{1+C_0 a^{-2n}}\right),
\ee
and its integration
\be
I\equiv\int a^{-1} \left(\alpha+\sqrt{1+C_0 a^{-2n}}\right)^{-\frac12}\,\dd a=\beta t.
\ee
Obviously the Chebyshev theorem is not applicable since it is not a binomial type. Nevertheless, we choose
\be
1+C_0 a^{-2n}=u^2\quad\mbox{so that } a=C_0^{\frac1{2n}}(u^2-1)^{-\frac1{2n}}.
\ee
Then we have
\be
I=-\frac1n\int u(u^2-1)^{-1}(\alpha +u)^{-\frac12}\,\dd u.
\ee
Finally we take $\alpha+u =w^2$. Then
\be
I=-\frac2n\int\frac{w^2-\alpha}{(w^2-\alpha)^2 -1}\,\dd w,
\ee
whose integration up to an additive integration constant is readily computed for all range of values of $\alpha$ to give us the results:
\be\label{7.11}
I=\left\{\begin{array}{lll}-\frac1n\left(\frac1{2\sqrt{1+\alpha}}\ln\left|\frac{w-\sqrt{1+\alpha}}{w+\sqrt{1+\alpha}}\right|+\frac1{\sqrt{1-\alpha}}\arctan\frac w{\sqrt{1-\alpha}}\right),&\quad
|\alpha|<1,\\
-\frac1n\left(\frac1{2\sqrt{2}}\ln\left|\frac{w-\sqrt{2}}{w+\sqrt{2}}\right|-\frac1w\right),&\quad \alpha=1,\\
-\frac1{2n}\left(\frac1{\sqrt{1+\alpha}}\ln\left|\frac{w-\sqrt{1+\alpha}}{w+\sqrt{1+\alpha}}\right|+\frac1{\sqrt{\alpha-1}}\ln\left|\frac{w-\sqrt{\alpha-1}}{w+\sqrt{\alpha-1}}\right|\right),&\quad \alpha>1,\\
-\frac1n\left(\frac1{\sqrt{2}}\arctan\frac w{\sqrt{2}}-\frac1w\right),&\quad \alpha=-1,\\
-\frac1n\left(\frac1{\sqrt{-(\alpha+1)}}\arctan\frac w{\sqrt{-(\alpha+1)}}+\frac1{\sqrt{1-\alpha}}\arctan\frac w{\sqrt{1-\alpha}}\right),&\quad \alpha<-1.
\end{array}\right.
\ee

Thus, in short, we conclude that the Born--Infeld fluid model (\ref{7.1}) allows an exact integration when $k=0$ for any value of the cosmological constant, although it lies beyond the
scope of the Chebyshev theorem.

We examine three simple, but interesting, cases.

\begin{enumerate}

\item[(i)] The cosmological constant vanishes, $\Lambda=0$. Hence $\alpha=-1$. From (\ref{7.11}) we read off the solution with the big-bang initial state
$a(0)=0$ or $w(0)=\infty$, which is given implicitly as follows,
\be
\frac1{nw}=\beta t+\frac1{\sqrt{2}n}\arctan\frac w{\sqrt{2}} -\frac\pi{2\sqrt{2}n},\quad t>0.
\ee
Consequently we obtain the power-growth law $a(t)\sim t^{\frac1n}$ as $t\to\infty$, which implies that the absence of a cosmological constant leads to
the absence of dark
energy.

\item[(ii)] The cosmological constant is large enough such that $\alpha>1$ or
\be\label{7.13}
\Lambda>16\pi G_n.
\ee
As before we see that the big bang solution with 
$
a(0)=0
$
is given by
\be
\frac1{2n}\left(\frac1{\sqrt{1+\alpha}}\ln\left|\frac{w-\sqrt{1+\alpha}}{w+\sqrt{1+\alpha}}\right|+\frac1{\sqrt{\alpha-1}}\ln\left|\frac{w-\sqrt{\alpha-1}}{w+\sqrt{\alpha-1}}\right|\right)+\beta t=0,
\ee
and
 enjoys the asymptotic behavior
\be\label{7.15}
a(t)\sim \left(\frac{C_0}{8(1+\alpha)}\right)^{\frac1{2n}}\e^{\beta\sqrt{1+\alpha} t}
=\left(\frac{\pi G_n C_0}{\Lambda}\right)^{\frac1{2n}}\e^{\sqrt{\frac{2\Lambda}{n(n-1)}}\,t},\quad t\to\infty.
\ee
It may be curious but not surprising to note that Newton's constant is absent from the dark energy because it is now ``overwhelmed" by the cosmological constant
realized as a significant background energy, as dictated by the condition (\ref{7.13}).

\item[(iii)] The cosmological constant is positive but small such that $|\alpha|<1$, or equivalently,
\be
0<\Lambda<16\pi G_n.
\ee
Then, inserting the big-bang initial condition $w(0)=\infty$ in (\ref{7.11}), we have
\bea\label{7.17}
&&\frac\pi{2n\sqrt{1-\alpha}}-\frac1n\left(\frac1{2\sqrt{1+\alpha}}\ln\left|\frac{w-\sqrt{1+\alpha}}{w+\sqrt{1+\alpha}}\right|+\frac1{\sqrt{1-\alpha}}\arctan\frac w{\sqrt{1-\alpha}}\right)\nn\\
&&=\beta t,\quad t>0.
\eea
From (\ref{7.17}) we see that the asymptotic behavior $a(\infty)=\infty$ corresponds to $w(\infty)=\sqrt{1+\alpha}$, which renders the estimate
\be\label{7.18}
a(t)\sim \e^{\lm t},\quad \lm=\sqrt{1+\alpha}\,\beta=\sqrt{\frac{2\Lambda}{n(n-1)}},\quad t\to\infty,
\ee
in which Newton's constant happens to be exactly wiped off, by virtue of (\ref{7.5}), which yields the {\em same} exponential growth rate as in (\ref{7.15}), independent of Newton's constant but only dependent on
the cosmological constant. This result,  (\ref{7.18}),  is parallel to (\ref{3.2}) for linear fluid models, for the absence of Newton's constant in the exponential growth rate.
These results are in sharp contrast with the universal formula (\ref{5.12}), for the Chaplygin fluid universe, which is explicitly related to Newton's constant.

\end{enumerate}

Conclusion: Although the Born--Infeld equation of state (\ref{7.1}) leads to a Chaplygin type ``fluid" with ``updated pressure $\tilde{P}_m$ and mass density
$\tilde{\rho}_m$",  as expressed
in (\ref{7.2}), it cannot lead to the presence of dark energy without a positive cosmological constant.

\end{document}